\documentclass{PoS}

\def\SoneG{$S_{1.4GHz}$}

\title{Space densities of radio AGN: The CoNFIG sample}

\ShortTitle{CoNFIG}

\author{\speaker{Melanie A. Gendre}\\
        Department of Physics and Astronomy, University of British Columbia, Vancouver, B.C., Canada V6T 1Z1\\
        E-mail: \email{mgendre@phas.ubc.ca}}

\author{Jasper V. Wall\\
        Department of Physics and Astronomy, University of British Columbia, Vancouver, B.C., Canada V6T 1Z1\\
        E-mail: \email{jvw@phas.ubc.ca}}

\abstract{The Combined NVSS-FIRST Galaxy (CoNFIG) survey
was defined by selecting all sources with \SoneG$\ge$1.3Jy from
the NRAO VLA Sky Survey (NVSS) in the north field of the Faint Images
of the Radio Sky at Twenty-cm (FIRST) survey. 
We carried out FRI/FRII morphology classification from NVSS and FIRST survey data;
to complete this process, new 8GHz VLA observations for 31
sources were obtained at 0.24 arcsec resolution. Optical identifications and redshift information 
were compiled for about
$\sim$80\% of the 270 radio sources in the sample, the mean
redshift being $\sim$0.6.

A major goal of this sample is a better definition of
the individual luminosity distributions and source counts for FRI and FRII sources,
 in order to determine accurately the evolution of the
luminosity function for both types. Amongst the aims are the issues of whether
the two populations are really distinct, whether physical evolution schemes permit one type to
evolve into the other, whether the dual-population unified scheme for radio AGN remains viable, and the
role radio AGN - star-formation feedback mechanisms.}

\FullConference{From planets to dark energy: the modern radio universe\\
		 October 1-5 2007\\
		 University of Manchester, Manchester, UK}

\begin{document}

The CoNFIG sample is defined as all sources with \SoneG $\ge$
1.3Jy from the NVSS catalogue within the north region of the FIRST
survey (2.95 sr defined roughly by $-8\deg \le dec \le 64\deg$ and $7
hrs \le ra \le 17 hrs$; Fig.~1). This selection includes resolved sources in
FIRST and NVSS with components having \SoneG $<$1.3Jy but with a total
flux density greater than 1.3Jy.\\

\begin{figure}[h]
  \begin{minipage}{7.5cm}
    \centerline{
      \includegraphics[angle=-90,scale=0.25]{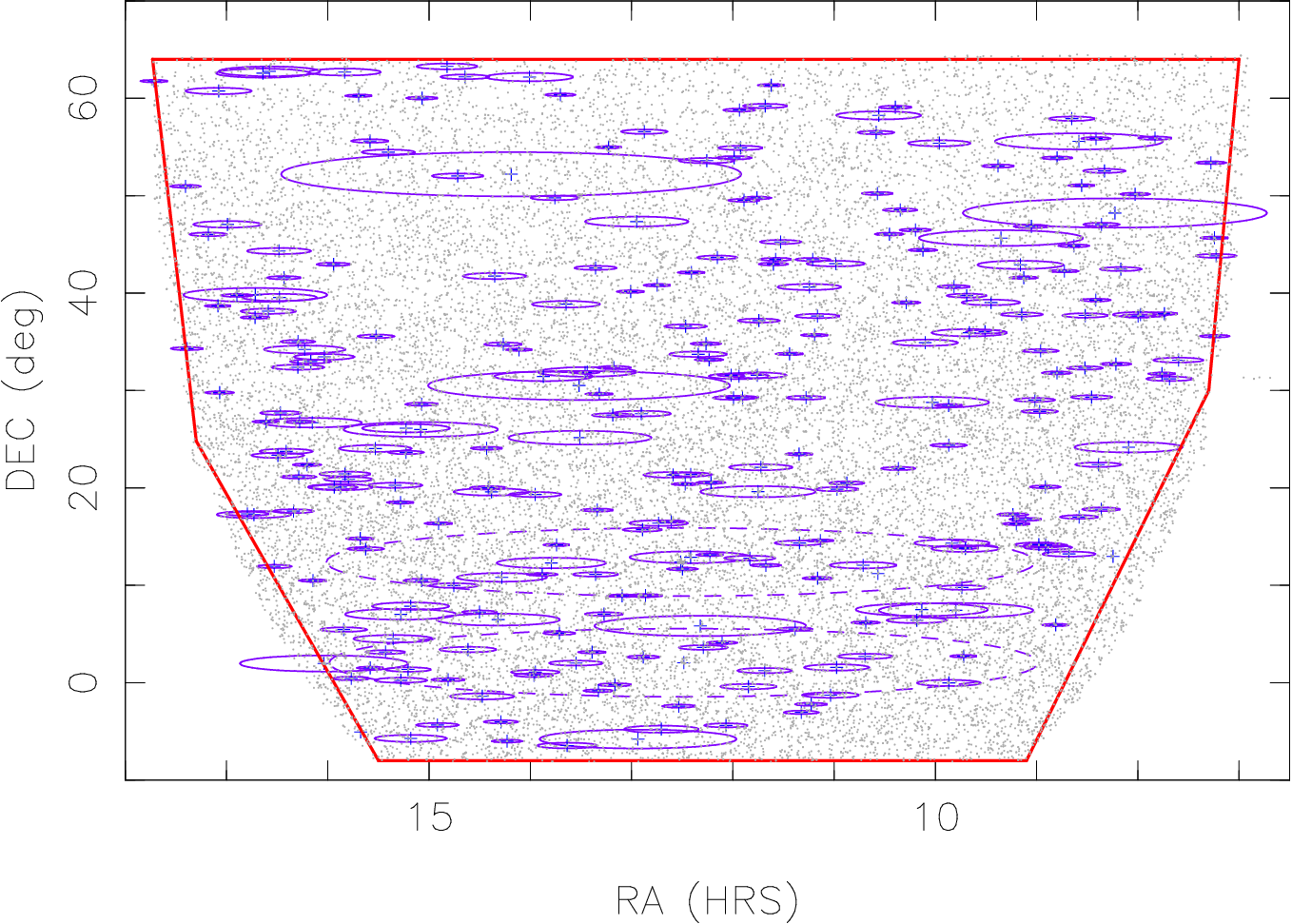}}
  \caption{The 270 sources in the CoNFIG sample. The grey area corresponds to the north
    region of the FIRST survey while the red contour delimits the
    region of the CoNFIG sample. Each source has
    \SoneG$\ge$1.3Jy and  is represented by
    a blue cross and circle. The radius of the
    circle is proportional to the flux of the source, with the
    exception of the two dashed circles corresponding to 3C\,273 (\SoneG=55Jy) and M87 (\SoneG=142Jy).}
  \end{minipage}
\hfill
  \begin{minipage}{6.0cm}
    \centerline{
      \includegraphics[angle=270,scale=0.35]{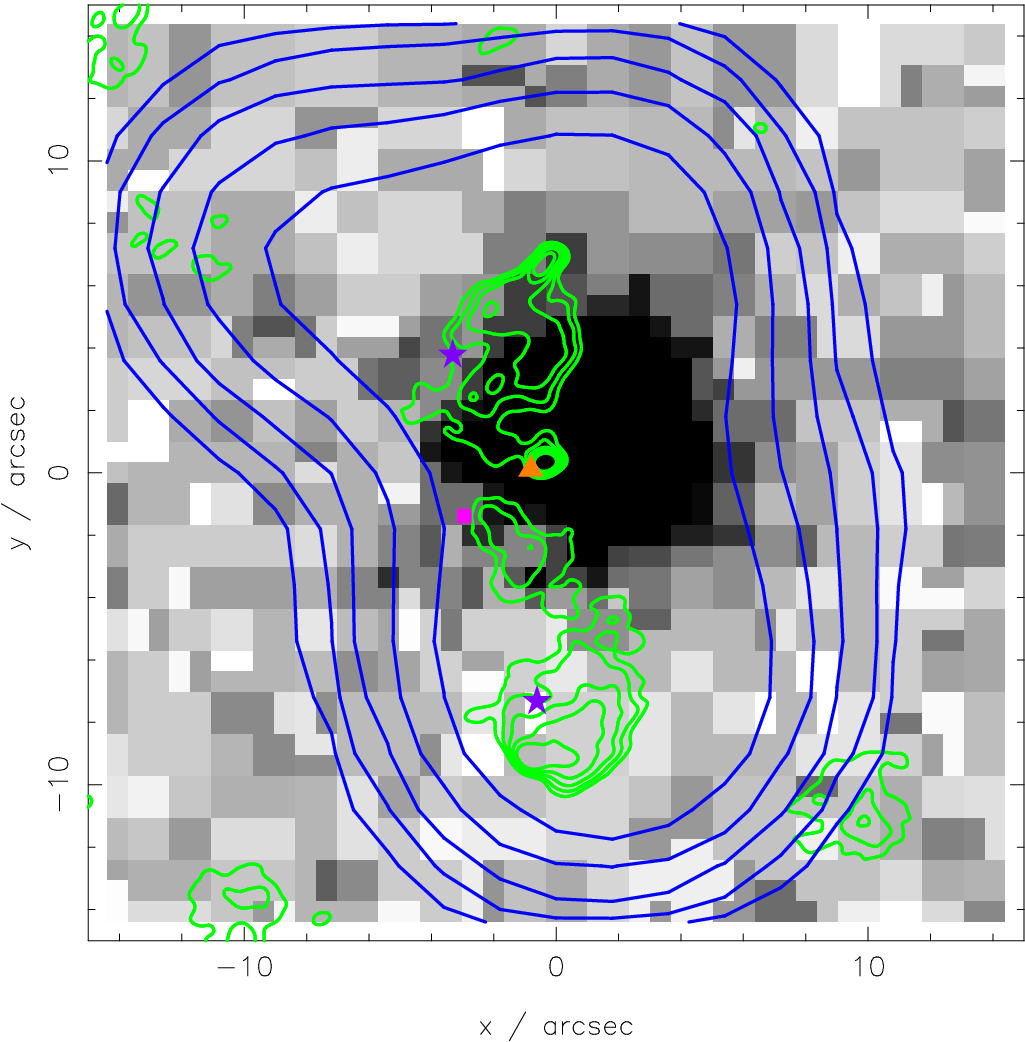}}
    \caption{An example of morphology typing. The blue contours are from the FIRST survey,
while the green contours are from our new 8GHz VLA observations, which show clearly
that the object is FRII in type -- the hot spots are at the extremities.}
  \end{minipage}
\end{figure}

The FRI/FRII morphology [1] of each source in 
the sample was determined by looking
at the FIRST and NVSS radio contour plot. If the contour plot showed
distinct hot spots at the edge of the lobes, and the lobes are
aligned, the source was classified as FRII. Sources with collimated
jets showing hot spots or jets close to the core were classified as
FRI. Most irregular looking sources were also classified as
FRI. Complementary VLA radio observation at 8GHz were carried out for
31 extended sources with uncertain morphology; an example is shown in Fig.~2. 
In this manner over 60 per cent of sources in
the CoNFIG sample were classified either as FRI or FRII.\\

Redshift information was retrieved for 220 sources ($\sim$80\% of the
sample). Redshifts range from z=0.0034 to3.522 with a median
redshift of z=0.6. Both FRI and
FRII redshift distributions peak at low redshifts (z $\le$ 0.5),
although the FRII redshift distribution covers a wider range, up to z=2.\\

Flux densities at different frequencies (178MHz, 365MHz, 408MHz, 2.7GHz and
5.0GHz) for each source were compiled to compute the
K-correction. The luminosities were then calculated.
Fig.~3 shows that the luminosity distribution of FRI
sources peaks at lower luminosities than FRIIs. This is no surprise as
FRIs are in general less powerful than FRIIs [1,2]. Compact sources on the
other hand are generally quasars of high luminosity.\\
\begin{figure}[h]
  \begin{minipage}{6.0cm}
  \centerline{
    \includegraphics[scale=0.3]{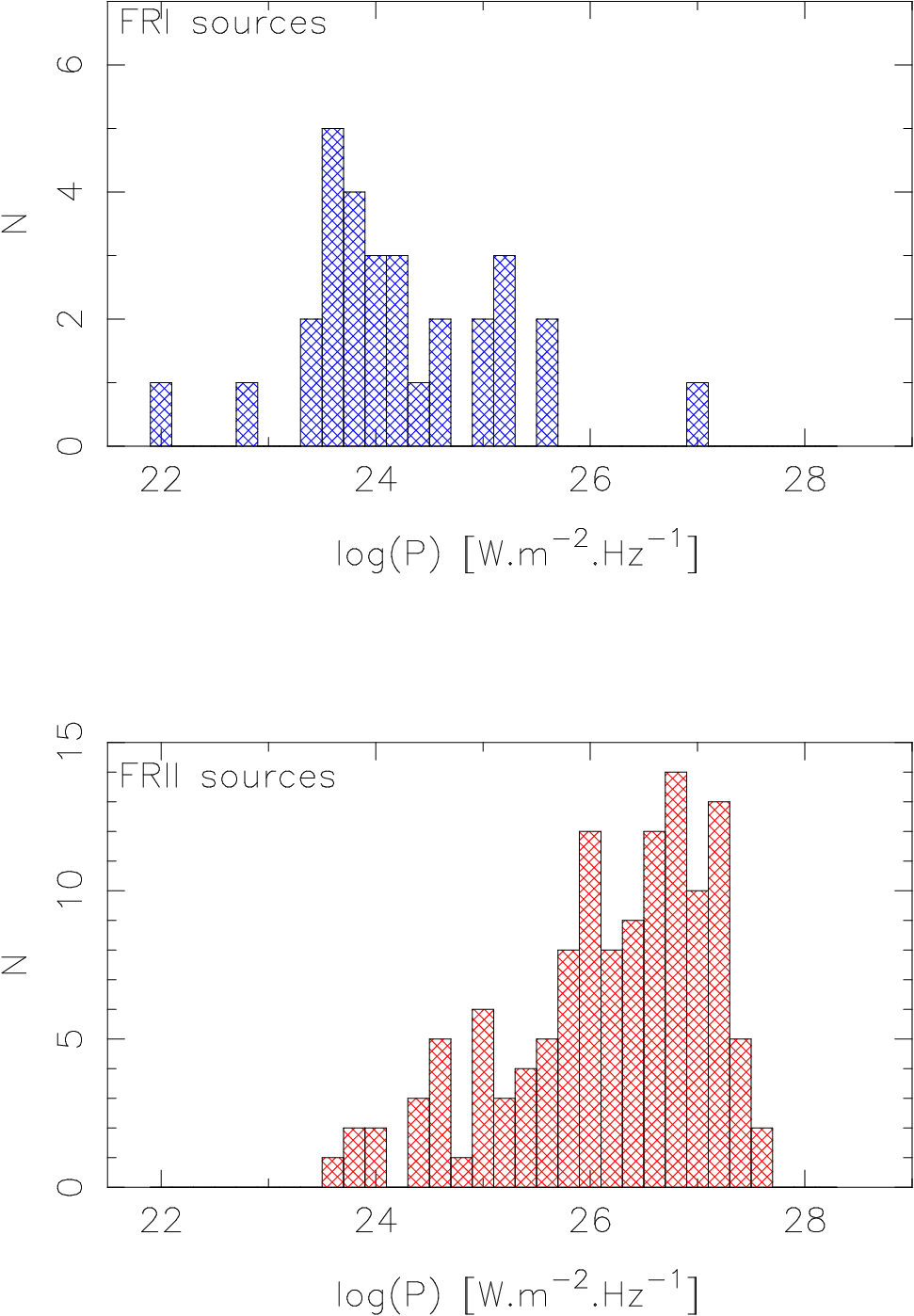}}
  \caption{Luminosity distribution by source
  type. FRI sources are generally less powerful than FRII
  sources; compact sources (generally quasars) have predominately high luminosities.}
\end{minipage}
\hfill
\begin{minipage}{7.0cm}
\centerline{
\includegraphics[scale=0.30,angle=-90]{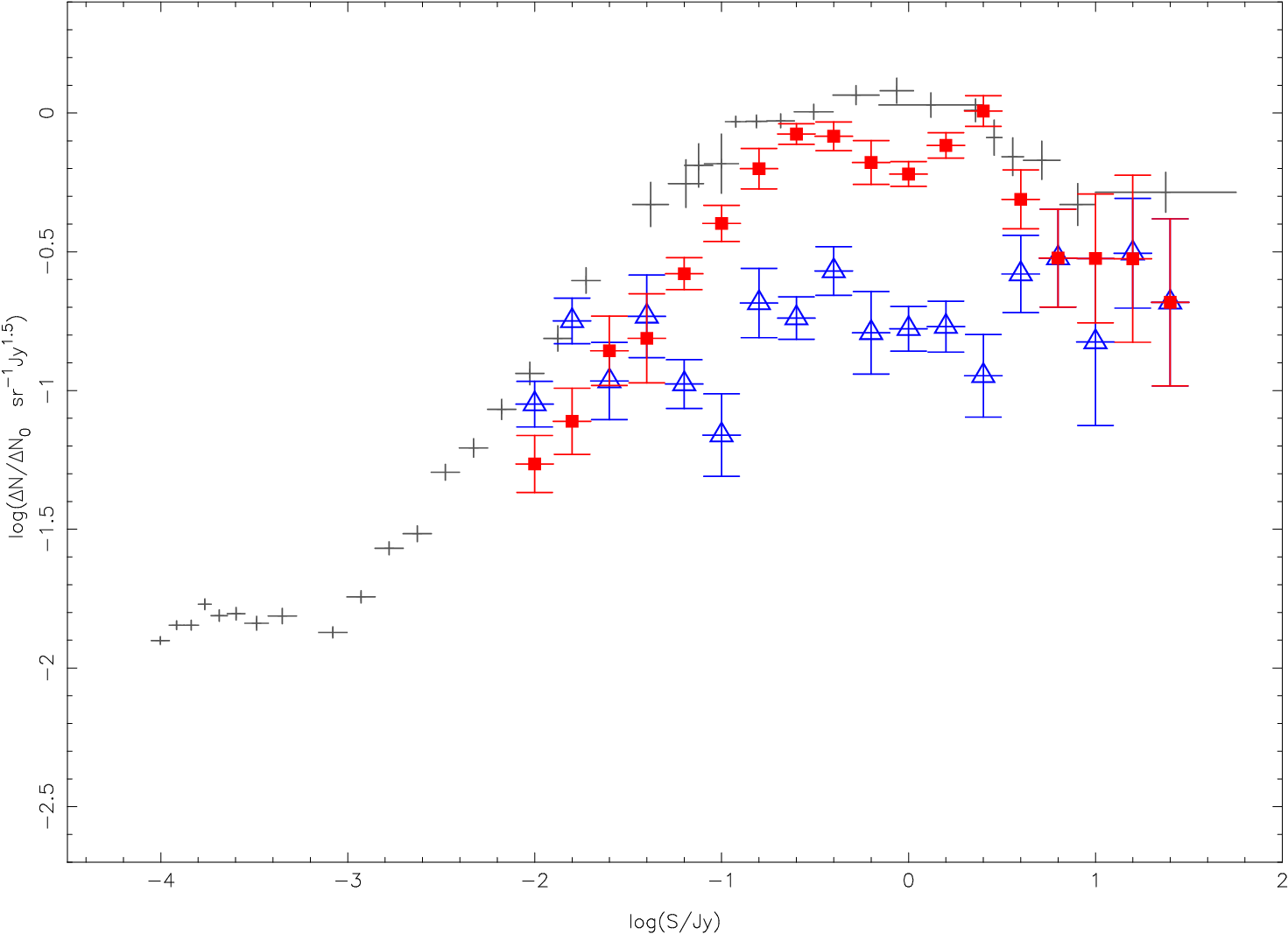}}
\caption{Relative differential source count per sr
  $\Delta N/\Delta N_0$ for  all sources (green), FRI (blue) sources, and FRII (red) sources. The NVSS source count is shown as grey crosses 
  for comparison. Here, $\Delta N_0 = 1200 \Delta
  (S^{-1.5})$ and the error bars correspond to $\sqrt{N}$ where N is
  the number of object in each bin. The FRII sources dominate the
  count down to the  lower flux densities, where FRI sources take over.}
  \end{minipage}
\end{figure}

In order to compute FRI and FRII source counts, several samples at
different flux limits (7.2 mJy, 50 mJy, 0.2 Jy, 0.8 Jy and 1.3 Jy)
were used. The morphology of the sources in each sample was determined
either by using the FIRST and NVSS contour plots, or, in the case of
the 7.2~mJy sample (the CENSORS sample [3]), by
using the Ledlow-Owen relation [2].\\

The relative differential source counts $\Delta N/\Delta N_0$ for the
total number of FRI and FRII sources
were then computed from the combined samples of $\sim$240 FRI and $\sim$340 FRII sources. 
As seen in Fig.~4, the FRII source count rises and
falls more rapidly than the FRI count, generally following the `evolution bulge'. In
contrast the FRI sources show a relatively flat count, 
exceeding the numbers of FRII sources only below log\SoneG$\le$-1.
Sadler et al. [4] showed that low-luminosity radio AGNs undergo mild evolution.
This first ever morphological source count agrees: it shows that the FRI sources clearly undergo
evolution, but much milder in form than that of the FRII sources.\\


\begin{thebibliography}{99}
\bibitem{Fanaroff} Fanaroff, B. \& Riley, J., 1974, MNRAS, 167:31 
\bibitem{Ledlow}Ledlow, M. J. \& Owen, F. N., 1996, ApJ, 112:9
\bibitem{Best} Best, P. N. et al., 2003, MNRAS, 346:627
\bibitem{Sadler}Sadler et al., 2007, MNRAS, 381:211

\end{thebibliography}
\end{document}